\documentclass[10pt,a4paper]{article}
\usepackage{amsmath,latexsym,amssymb,amsfonts}
\usepackage[dvips]{graphicx}
\usepackage{bm}
\allowdisplaybreaks

\begin{document}

\title{\textbf{No-boundary measure and preference for\\large $e$-foldings in multi-field inflation}}
\author{\textsc{Dong-il Hwang}$^{a,b}$\footnote{dongil.j.hwang@gmail.com},\;\;
\textsc{Soo A Kim}$^{c}$\footnote{sooastar@gmail.com},\;\;
\textsc{Bum-Hoon Lee}$^{a}$\footnote{bhl@sogang.ac.kr},\;\;\\
\textsc{Hanno Sahlmann}$^{c,d}$\footnote{hanno.sahlmann@gravity.fau.de}\;\;
and \textsc{Dong-han Yeom}$^{a,e}$\footnote{innocent.yeom@gmail.com}\\
\textit{$^{a}$\small{Center for Quantum Spacetime, Sogang University, Seoul 121-742, Republic of Korea}}\\
\textit{$^{b}$\small{Research Institute for Basic Science, Sogang University, Seoul 121-742, Republic of Korea}}\\
\textit{$^{c}$\small{Asia Pacific Center for Theoretical Physics, Pohang 790-784, Republic of Korea}}\\
\textit{$^{d}$\small{Department of Physics, University Erlangen-Nueremberg, 91058Erlangen, Germany}}\\
\textit{$^{e}$\small{Yukawa Institute for Theoretical Physics, Kyoto University, Kyoto 606-8502, Japan}}
}
\maketitle

\begin{flushright}
{\tt  APCTP-Pre2012-009\\
YITP-13-58}
\end{flushright}

\begin{abstract}
The no-boundary wave function of quantum gravity usually assigns only very small probability to long periods of inflation. This was a reason to doubt about the no-boundary wave function to explain the observational universe. We study the no-boundary proposal in the context of multi-field inflation to see whether the number of fields changes the situation. For a simple model, we find that indeed the no-boundary wave function can give higher probability for sufficient inflation, but the number of fields involved $N_{f}$ has to be very high, e.g., $N_{f} \simeq m^{-2}$.
\end{abstract}

%\newpage
%
%\tableofcontents
%
\newpage

\section{Introduction}
\label{sec_Intro}

The nature of the initial singularity of our universe is an important but unsolved problem in modern cosmology.
The traditional approach to this problem is to canonically quantize the universe and study the wave function, by solving the Wheeler-DeWitt equation \cite{DeWitt:1967yk}. There is no general consensus on the boundary condition of the wave function for our universe. However, perhaps one natural way to address this problem is to consider that our universe is in the ground state. Hartle and Hawking \cite{Hartle:1983ai} observed that the ground state can be written in terms of the Euclidean path integral. Thus they suggested an analogous extension to cosmology including gravitation, so-called the no-boundary wave function.

The no-boundary wave function can be interpreted as a superposition of space-time histories. It can be approximated by sum-over on-shell solutions, so-called instantons. Traditionally, there are two approaches to calculate the instantons. One way is following the technique of \cite{Hartle:1983ai}: by assuming the scalar field to be approximately constant, integrate the Euclidean action over a part of the four-sphere metric \cite{Kiefer}. This result exponentially prefers small vacuum energy, and hence some people believed that our universe probably has zero vacuum energy \cite{Hawking:1984hk}. The other way is to obtain the wave function by solving the Wheeler-DeWitt equation directly \cite{Vilenkin:1994rn}. In this approach, again, we have to assume the slow-rolling limit; hence, we can reduce the equation as an ordinary differential equation. One strong point of this approach is that this approach allows some alternative boundary conditions (e.g., \cite{Vilenkin:1986cy}). Again, the Hartle-Hawking wave function exponentially prefers small $e$-foldings.

All these approaches rely on analytic techniques and had to assume the slow-roll limit. However, in general situations, we have to consider dynamical instantons. Note that the sum-over instantons can have problems of divergence \cite{Vilenkin:1994rn}, which may be alleviated by considering complex valued instantons \cite{Halliwell:1989dy}, so-called \textit{fuzzy instantons} \cite{Lyons:1992ua, Hartle:2007gi, Hartle:2008ng}. If we use fuzzy instantons, by using numerical techniques, we can deal with the dynamical -- even fast-rolling -- situations.

Although the instantons should be complexified, we see a real valued world; in other words, we see a classical universe. This means that not all the complex-valued instantons are allowed, but only certain special instantons that approach to real valued functions are indeed allowed. For this preferred subset of histories, the no-boundary wave function assigns a \emph{semi-classical} probability distribution, so-called the no-boundary measure. Technically, to find classicalized histories, we need a searching process. This was studied analytically and approximately by Lyons \cite{Lyons:1992ua} and numerically and more precisely by Hartle, Hawking and Hertog \cite{Hartle:2007gi,Hartle:2008ng}. In \cite{Hartle:2007gi,Hartle:2008ng}, they calculated for a quadratic potential that has only one local minimum in Einstein gravity; the present authors studied the no-boundary measure for more general examples: scalar-tensor gravity \cite{Hwang:2011mp} and fast-rolling fields around the hill-top \cite{Hwang:2012zj}. In this paper, we will use the same numerical algorithm that was developed by the present authors.

Due to previous studies, it was known that the no-boundary measure disfavors the histories which have a large number of $e$-foldings. If we believe that the early universe had experienced more than $60$ $e$-foldings from primordial inflation, the no-boundary measure is not compatible with this expectation.

On these grounds, one may reject the no-boundary wave function in favor of other prescriptions, for example Vilenkin’s tunneling proposal \cite{Vilenkin:1986cy}, which may however be regarded as less natural on theoretical grounds. Perhaps, it can be better to obtain large $e$-foldings, but the out-going mode on the superspace needs further explanations; moreover, if the potential is sufficiently steep and hence the field becomes dynamic, then it is not clear how to calculate the wave function precisely. On the other hand, by accepting the fact that the no-boundary wave function disfavor large $e$-foldings, but
\begin{itemize}
\item try to avoid the need for a long period of inflation such as ekpyrotic universe \cite{Khoury:2001wf}, string gas cosmology \cite{Brandenberger:1988aj}, the big bounce model \cite{Khoury:2001bz}, or the pre-big-bang model \cite{Gasperini:1992em},

\item justify inflation on anthropic grounds, for example in the context of the multiverse \cite{Susskind:2003kw},

\item lift up the inflationary probabilities by considering the volume weighting \cite{Hawking:2002af,Hawking:2006ur}.\end{itemize}
These proposals each come with their own questions and difficulties attached. Therefore in this article we explore another mechanism to lift up the probability for the larger $e$-foldings in the context of the no-boundary measure.
We study the no-boundary measure of \textit{multiple fields} and argue that \textit{the degeneracy of the large number of fields can lift up the probability of inflationary histories} without considering any additional factors.
Although the Euclidean action does not prefer inflationary histories, there exist many numbers of histories which produce larger $e$-foldings, if we have approximately $m^{-2}$ number of fields, where $m$ is the mass of each field.

In Section~\ref{nb_proposal}, we briefly review the no-boundary wave function and explain the reason why it does not prefer inflationary histories.
In Section~\ref{sec_nb_Nflation}, the no-boundary measure is extended to the multi-field case and $N$-flation models, and it is demonstrated that the degeneracy can lift up the probability of larger $e$-foldings.
We also relate the model with the observations. Finally, we summarize the results and discuss possible extensions in Section~\ref{sec_Conclusion}.

\section{No-boundary wave function}
\label{nb_proposal}
The ground state of the universe, or the no-boundary condition of the wave function of the universe, can be described by the Euclidean path integral \cite{Hartle:1983ai}:
\begin{eqnarray}
\Psi[h_{\mu\nu}, \chi] = \sum_\mathcal{T} \int \mathcal{D}g_{\mu\nu} \mathcal{D} \phi \; e^{-S_{\text{E}}[g_{\mu\nu}, \phi]}, \label{nb_single_waveftn}
\end{eqnarray}
where $\mathcal{T}$ denote compact manifold topologies with a single spherical boundary, and
the metric $g_{\mu\nu}$ and the matter field $\phi$ take the value $h_{\mu\nu}$ and $\chi$ at the boundary. Here, $S_{\text{E}}$ is the Euclidean action: we assume Einstein gravity and a minimally coupled scalar field,
\begin{eqnarray}
S_{\text{E}} = - \int d^{4}x \sqrt{+g} \left( \frac{1}{16\pi} R - \frac{1}{2} (\nabla \phi)^{2} - V(\phi) \right)
\end{eqnarray}
in the units $\hbar=G=c=1$.
In this paper, for simplicity, we restrict our interests to a minisuperspace; for a homogeneous and isotropic universe, the metric becomes
\begin{eqnarray}
ds^{2} = N(\lambda)^{2} d\lambda^{2} + a(\lambda)^{2} d\Omega_{3}^{2}.
\end{eqnarray}
In this context, it is convenient to view the time integral in the action as a contour integral in the complex time plane (for details see for example \cite{Hartle:2007gi}) and shift the contour in suitable ways. Thus, $\lambda$ (which we take to be in $[0,1]$) is a parametrization of the chosen contour.
The no-boundary wave function takes the form
\begin{eqnarray}
\Psi[b,\chi] = \int \mathcal{D}N \mathcal{D}a \mathcal{D}\phi \; e^{-S_{\text{E}}[a,\phi]},
\end{eqnarray}
where $a(\lambda=1)=b$ and $\phi(\lambda=1)=\chi$ are the boundary values at the only boundary of space-time, and at $\lambda=0$ one imposes conditions suitable for a regular metric (``no-boundary conditions").

The path integral can be approximated by the method of steepest descent. To lowest order, in this approximation we only keep the contribution from saddle points of the exponent:
\begin{eqnarray}
\label{eq_saddle}
\Psi[b,\chi] \simeq \sum_p e^{-S_{\text{E}}^p[b,\chi]}
\end{eqnarray}
where $S_{\text{E}}^p[b,\chi]$ is the action for the solution labeled by $p$ with the boundary conditions
$b,\chi$, of the equations
\begin{equation}
\ddot{\phi} = - 3 \frac{\dot{a}}{a} \dot{\phi} + \frac{\partial V(\phi)}{\partial \phi},\qquad \qquad
\ddot{a} = - \frac{8 \pi}{3} a \left(\dot{\phi}^{2} + V(\phi) \right),
\label{eq_single}
\end{equation}
where `$\dot{~}$' denotes derivative with respect to $\tau$ defined by
\begin{eqnarray}
\tau(\lambda) \equiv \int^{\lambda} d\lambda' N(\lambda').
\end{eqnarray}
For the detailed and convenient choice of a time contour, see the contour diagram in Ref.~\cite{Lyons:1992ua}. The on-shell action can be simplified to
\begin{eqnarray}
S_{\text{E}} = 4\pi^{2} \int d \tau \left( a^{3} V - \frac{3}{8 \pi} a \right).
\end{eqnarray}

We are especially interested in regions in which the wave function behaves almost classically. In the saddle point approximation, Equation~\eqref{eq_saddle}, this requires the condition so-called the classicality,
\begin{equation}
|\nabla S_{\text{E}}^{\text{Re}}[b,\chi]| \ll |\nabla S_{\text{E}}^{\text{Im}}[b,\chi]|, \label{eqn:classicality}
\end{equation}
where $S_{\text{E}}^{\text{Re}}$ and $S_{\text{E}}^{\text{Im}}$ are the real and imaginary parts of the Euclidean action and $\nabla$ is a derivation that is defined on the superspace. In other words, among various complex valued on-shell histories, we should select some of them that satisfy this condition. To select such histories, in \cite{Hwang:2011mp}, we discussed the details of the numerical searching algorithm; the same algorithm is adopted in this paper.
We call them \emph{classicalized histories}. The probability for classicalized histories becomes
\begin{eqnarray}
dP[b,\chi] \varpropto |\Psi[b,\chi]|^{2}\, db\,d\chi \simeq e^{-2 S_{\text{E}}^{\text{Re}}[b,\chi]}\, db\,d\chi. \label{classicality}
\end{eqnarray}

The no-boundary measure is defined on the boundary of the histories, $b$ and $\chi$.
However, it is convenient to label a history not by the boundary values, but by the initial conditions, for slow-roll cases. Moreover, if the field slowly moves, then the Euclidean action is well approximated by the potential energy at the turning point from the Euclidean time to the Lorentzian time \cite{Hwang:2012mf}:
\begin{eqnarray}
S_{\text{E}} \simeq - \frac{3}{16} \frac{1}{V(\varphi)},
\end{eqnarray}
where $\varphi \equiv \phi(t=0)$. In these cases, one thus has
\begin{eqnarray}
\label{nb_prob_Single_Quad}
dP[\varphi] \varpropto \exp \left(\frac{3}{8} \frac{1}{V(\varphi)}\right)\, d\varphi.
\end{eqnarray}
as an approximation of the probability to observe a history starting from $\varphi$ in the Euclidean regime.

We take the simplest inflation model for single field as one having a quadratic potential \cite{Linde83}
\begin{eqnarray}
V(\phi) = \frac{1}{2} m^{2} \phi^{2} \label{quadpotential}.
\end{eqnarray}
In \cite{Hartle:2008ng}, Hartle, Hawking and Hertog could show by numerically and analytically such that, except for $\varphi = 0$, only $\varphi$ larger than a critical value has the solution space satisfying the classicality: Equation~\eqref{eqn:classicality}. That is, the classicalized history exists only for
\begin{eqnarray}
\varphi > \varphi^{c} \sim 0.62. \label{phi_cutoff}
\end{eqnarray}
One may worry about the point $\varphi = 0$ since the action as well as the no-boundary measure could diverge there. However, due to the divergence it is impossible to evaluate the derivatives in the classicality condition, Equation~\eqref{eqn:classicality}. Furthermore, $\varphi = 0$ is surrounded by non-classical points in field space, and it is hard to see how the action would even be defined as a differentiable function. Therefore, it is reasonable to regard $\varphi = 0$ as a non-classical point and it is not considered in the followings.

In the slow-roll approximation~\cite{slowroll}, the total $e$-folding number obtained from $\varphi$ is approximated by
\begin{eqnarray}
\mathcal{N}_{e} \simeq - 8 \pi \int_{\varphi} d \phi \; \frac{V(\phi)}{V(\phi)_{,\phi}}.
\label{Ne}
\end{eqnarray}
Of course, when the scalar field rolls around the local minimum after the inflation, few $e$-foldings can be added, but we can ignore them. The previous formula can be further reduced for the quadratic potential Equation~\eqref{quadpotential}:
\begin{eqnarray}
\mathcal{N}_{e} \simeq 2 \pi \varphi^{2}. \label{formula_Ne}
\end{eqnarray}
Then, the no-boundary measure, Equation~\eqref{nb_prob_Single_Quad}, can be expressed as a function of $e$-folding number\footnote{Since the expansion during the Euclidean time is negligible, Equation~\eqref{formula_Ne} is still a good approximation.}
\begin{eqnarray}
dP[\mathcal{N}_{e}] \varpropto \exp \left(\frac{3 \pi}{2} \frac{1}{m^{2} \mathcal{N}_{e}}\right)\, d\mathcal{N}_e. \label{nb_single_Ne}
\end{eqnarray}
Here, the Jacobian of the change of variables is ignored because the exponential part is dominant.
The classicalized histories exist for
\begin{eqnarray}
\mathcal{N}_{e} > \mathcal{N}_{e}^{c} \equiv 2 \pi (\varphi^{c})^{2} \sim 2.4.
\label{eq_lowNcutoff}
\end{eqnarray}
Therefore, the no-boundary measure exponentially prefers a small $e$-folding number ($\sim 2.4$).

As discussed in Section~\ref{sec_Intro}, there are several possible interpretations of this disagreement between the prediction of the no-boundary measure and the inflationary universe.
One of the ideas is the volume weighting. It enhances the likelihood of larger $e$-foldings by multiplying the volume ($\equiv \exp 3\mathcal{N}_{e}$) for each history \cite{Hartle:2007gi},
\begin{eqnarray}
dP[\mathcal{N}_{e}] \varpropto \exp \left(\frac{3 \pi}{2} \frac{1}{m^{2} \mathcal{N}_{e}} + 3 \mathcal{N}_{e}\right)\, d\mathcal{N}_e.
\end{eqnarray}
This volume weighting sufficiently lifts up the probability of larger $e$-foldings, if
\begin{eqnarray}
\mathcal{N}_{e} \gtrsim m^{-2} \label{cri_Ne}.
\end{eqnarray}
In the following, we will study under which conditions the addition of further fields may have a similar effect.

\section{No-boundary measure in multi-field inflation}
\label{sec_nb_Nflation}
In this section, we study the no-boundary measure with multiple fields. As simple toy model, we take all fields, having same quadratic potential type for simplicity as in the earlier works on $N$-flation models~\cite{DKMW,KL061}\footnote{Earlier investigations of the $N$-flation paradigm~\cite{DKMW,KL061,KL062,RM} have worked under the assumption that all relevant fields are close to their minima and can be described by quadratic potentials. For axions the full potential is trigonometric and recently Refs.~\cite{KLS} showed that the quadratic approximation is unreliable. The reason we follow the earlier works rather than the more recent ones is mainly for simplicity. We intend to do further work on for the case of more complicated potentials.} where it was shown that any of these fields does not need to take on values in excess of the Planck scale\footnote{Note that, in literature, people use \emph{reduced} Planck mass scale.} and a large number of fields, predicted by string vacuum solutions, is needed to get sufficient $e$-foldings.
We take $N$-flation models as a possible candidate of the multi-field inflation models throughout this paper.
The basic effect that we will be investigating is that, although large amounts of inflation are still suppressed in terms of the action, there are now many histories that contribute to a given amount of inflation. This volume in field space competes with the suppression due to the action.
\subsection{No-boundary wave function of multiple fields}
The no-boundary wave function in Equation~\eqref{nb_single_waveftn} can be extended to the multi-field case.
In the minisuperspace of scale factor and a set of scalar fields, it becomes
\begin{eqnarray}
\Psi[b,\vec{\chi}] = \int\mathcal{D}N \mathcal{D}a \mathcal{D}\vec{\phi} \; e^{-S_{\text{E}}[a,\vec{\phi}]} \label{multifield_waveftn}
\end{eqnarray}
where the arrow denotes multiple fields, $\vec{\phi} \equiv \{\phi_{j}\}$, and $a(\lambda=1)=b$ and $\vec{\phi}(\lambda=1)=\vec{\chi}$ and the parametrization $\lambda$ was chosen such that $\lambda=1$ corresponds to the single boundary.

As discussed in Section~\ref{nb_proposal}, we use the method of the steepest descent
\begin{eqnarray}
\Psi[b,\vec{\chi}] \simeq \sum_{p} e^{-S_{\text{E}}^{p}[b,\vec{\chi}]},
\end{eqnarray}
where
\begin{align}
\ddot{\phi_{i}} = - 3 \frac{\dot{a}}{a} \dot{\phi_{i}} + \frac{\partial V(\vec{\phi})}{\partial \phi_{i}},
\qquad\qquad
\ddot{a} = - \frac{8 \pi}{3} a \left(\dot{\vec{\phi}}^{2} + V(\vec{\phi}) \right),\\
S_{\text{E}} = 4\pi^{2} \int d \tau \left( a^{3} V(\vec{\phi}) - \frac{3}{8 \pi} a \right),
\end{align}
where the subscript $i$ denotes the $i$-th field.

\subsection{Degeneracy of solution space for the case of a quadratic potential}
We consider a set of uncoupled fields in quadratic potentials with the same mass $m_i = m$~\cite{DKMW,KL061},
\begin{eqnarray}
V(\vec{\phi}) = \frac{1}{2} m^{2} \vec{\phi}^{2}.
\end{eqnarray}
The equations of motion are
\begin{align}
\ddot{\vec{\phi}} = - 3 \frac{\dot{a}}{a} \dot{\vec{\phi}} + m^{2} \vec{\phi},\qquad\qquad
\ddot{a} = - \frac{8 \pi}{3} a \left(\dot{\vec{\phi}}^{2} + \frac{1}{2} m^{2} \vec{\phi}^{2} \right).
\end{align}
The no-boundary conditions at $\tau=0$ require $\dot{\vec{\phi}}=0$. Thus we can make the ansatz
\begin{equation}
\label{eq_ansatz}
\vec{\phi}(\tau)=\phi(\tau)\hat{\phi}, \qquad\qquad (\hat{\phi})^2=1,
\end{equation}
which requires $\dot{\vec{\phi}}=(\dot{\phi}/\phi)\vec{\phi}$. Then the equations of motion for $\phi(\tau)$ and $a(\tau)$ are precisely given by the Equations~\eqref{eq_single} for a single scalar in a quadratic potential.
The no-boundary wave function is thus independent of the direction of $\vec{\phi}$ in field space,
\begin{equation}
\Psi(b,\vec{\chi})\equiv \Psi(b,\chi), \qquad\qquad \chi=|\vec{\phi}|.
\end{equation}
As a result, the classicality conditions in this case reduce to those of the single field case, and one can use Equations~\eqref{nb_prob_Single_Quad} and \eqref{phi_cutoff}.
Similar to Equation~\eqref{nb_prob_Single_Quad}, the no-boundary measure can be defined on the initial field space in the slow-roll limit,
\begin{eqnarray}
dP[\vec{\varphi}]
	\varpropto \exp \left(\frac{3}{4} \frac{1}{m^{2} \varphi^{2}}\right) \, d\vec{\varphi} \label{nb_vec_phi}
\end{eqnarray}
and we have again ignored a Jacobian due to the change of variables.
 $\varphi\equiv |\vec{\varphi}|$.
Important physical observables, e.g., the $e$-folding number and power spectrum, do not depend on the angular part of $\vec{\varphi}$. Therefore, for these observables the effective probability distribution is
\begin{equation}
dP[\varphi]=\int d\hat{\varphi}\, dP(\hat{\varphi},\varphi)  \simeq D(\varphi^2) \exp \left(\frac{3}{4} \frac{1}{m^{2} \varphi^{2}}\right) \, d\varphi
\label{D_nb_phisq}
\end{equation}
where $|\hat{\varphi}|=1$ and $D(\varphi^2)$ is the surface area of the shell $|\vec{\varphi}|=\varphi$ in field space.

Now we need to discuss the issue of cutoffs in Equation~\eqref{D_nb_phisq}. For a single field model with the quadratic potential, to obtain sufficient $e$-foldings, we need super-Planckian field values, where this is not quite natural in terms of quantum field theoretical sense. One of the motivations of multi-field inflation is to avoid this problem \cite{DKMW,KL061,KL062,KLS}. This is the reason why we introduce cutoffs for each fields. (Later, we will comment that our main assertion -- a large number of fields can enhance large $e$-foldings -- does not depend on the cutoff structure.) Here, we assume that all fields are below $\alpha$ that is order of the Planck mass $\sim 1$.

Now we account $D(\varphi^2)$ from the cutoffs. To estimate $D(\varphi^2)$, we choose $\varphi_{i}$ randomly and count the number of resulting $\varphi^{2}$.
It can be done by assuming $\varphi_{i}$ as a uniform random variable and using the central limit theorem.
We assume that there are $N_{f}$ number of fields with initial field value $\varphi_{i}$ following
\begin{eqnarray}
\varphi_{i} \in \mathcal{U}\left(0, \alpha\right), \label{dist_init_phi}
\end{eqnarray}
where $\mathcal{U}(p, q)$ is the uniform distribution with minimum $p$ and maximum $q$.
For the uniform distribution of $\varphi_{i}$, the mean and the variance of $\varphi_{i}^{2}$ become
\begin{eqnarray}
E[\varphi_{i}^{2}] &=& \frac{1}{\alpha} \int_{0}^{\alpha} d\varphi_{i} \; \varphi_{i}^{2}
= \frac{\alpha^{2}}{3},\\
\sigma^{2}[\varphi_{i}^{2}] &=& E[\varphi_{i}^{4}] - E[\varphi_{i}^{2}]^{2}
= \frac{4 \alpha^{4}}{45}.
\end{eqnarray}
When $N_{f}$ is large, a statistical behavior will arise, e.g., it will be more probable to observe $\varphi^{2}$ around its expectation value.
It is described by the central limit theorem which states that the distribution of $E[\varphi^{2}]$ follows the Gaussian distribution $\mathcal{N}\left(\mu, \sigma^{2}\right)$ with the mean $\mu$ and the variance $\sigma^{2}$, such that
\begin{eqnarray}
E[\varphi^{2}] &\in& \mathcal{N}\left(\mu, \sigma^{2}\right),\\
\mu &\equiv& \frac{\alpha^{2} N_{f}}{3},\\
\sigma^{2} &\equiv& \frac{4 \alpha^{4} N_{f}}{45}.
\end{eqnarray}
This distribution corresponds to the relative number of equivalent choices of $\vec{\varphi}$ for a given $\varphi^{2}$.
Therefore,
\begin{eqnarray}
D(\varphi^{2}) \varpropto \exp \left(- \frac{\left(\varphi^{2} - \mu\right)^{2}}{2 \sigma^{2}}\right).
\end{eqnarray}
Then, the no-boundary measure of $\varphi^{2}$ follows from Equation~\eqref{D_nb_phisq}:
\begin{eqnarray}
P[\varphi^{2}] \varpropto \exp \left(\frac{3}{4} \frac{1}{m^{2} \varphi^{2}}
- \frac{\left(\varphi^{2} - \mu\right)^{2}}{2 \sigma^{2}}\right). \label{nb_multi_phisq}
\end{eqnarray}
This summarizes the Euclidean probability measure and the degeneracy of the configuration space.

\begin{figure}
\begin{center}
\includegraphics[scale=0.9]{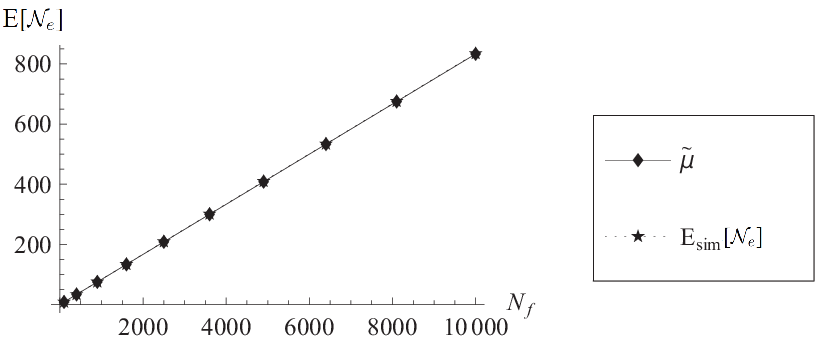}\nolinebreak
\includegraphics[scale=0.9]{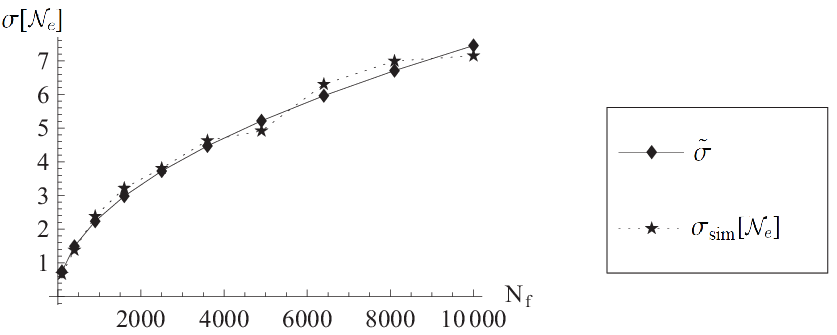}
\caption{\label{formula_vs_sim}Comparison between $\tilde{\mu}$ and $E_{\text{sim}}[\mathcal{N}_{e}]$ and between $\tilde{\sigma}^{2}$ and $\sigma_{\text{sim}}^{2}[{\mathcal{N}_{e}}]$. Subscript $\text{sim}$ denotes the numerical values obtained from $100$ simulations for each $N_{f}$. We used the uniform random $\varphi_{i}$ following Equation~\eqref{dist_init_phi} with $\alpha = 1/\sqrt{8\pi}$ which corresponds to the initial field value below the reduced Planck mass. Only Lorentzian evolution is considered in these simulations, since the expansion during the Euclidean time is negligible.}
\end{center}
\end{figure}
Now we turn our attention to the $e$-folding number.
The degeneracy factor can be described in terms of the $e$-folding number by using Equation~\eqref{formula_Ne},
\begin{eqnarray}
E[\mathcal{N}_{e}] &\in& \mathcal{N}\left(\tilde{\mu}, \tilde{\sigma}^{2}\right),\\
\tilde{\mu} &\equiv& \frac{2 \pi \alpha^{2} N_{f}}{3},\\
\tilde{\sigma}^{2} &\equiv& \frac{16 \pi^{2} \alpha^{4} N_{f}}{45}.
\end{eqnarray}

\begin{figure}[t]
\begin{center}
\includegraphics[scale=0.5]{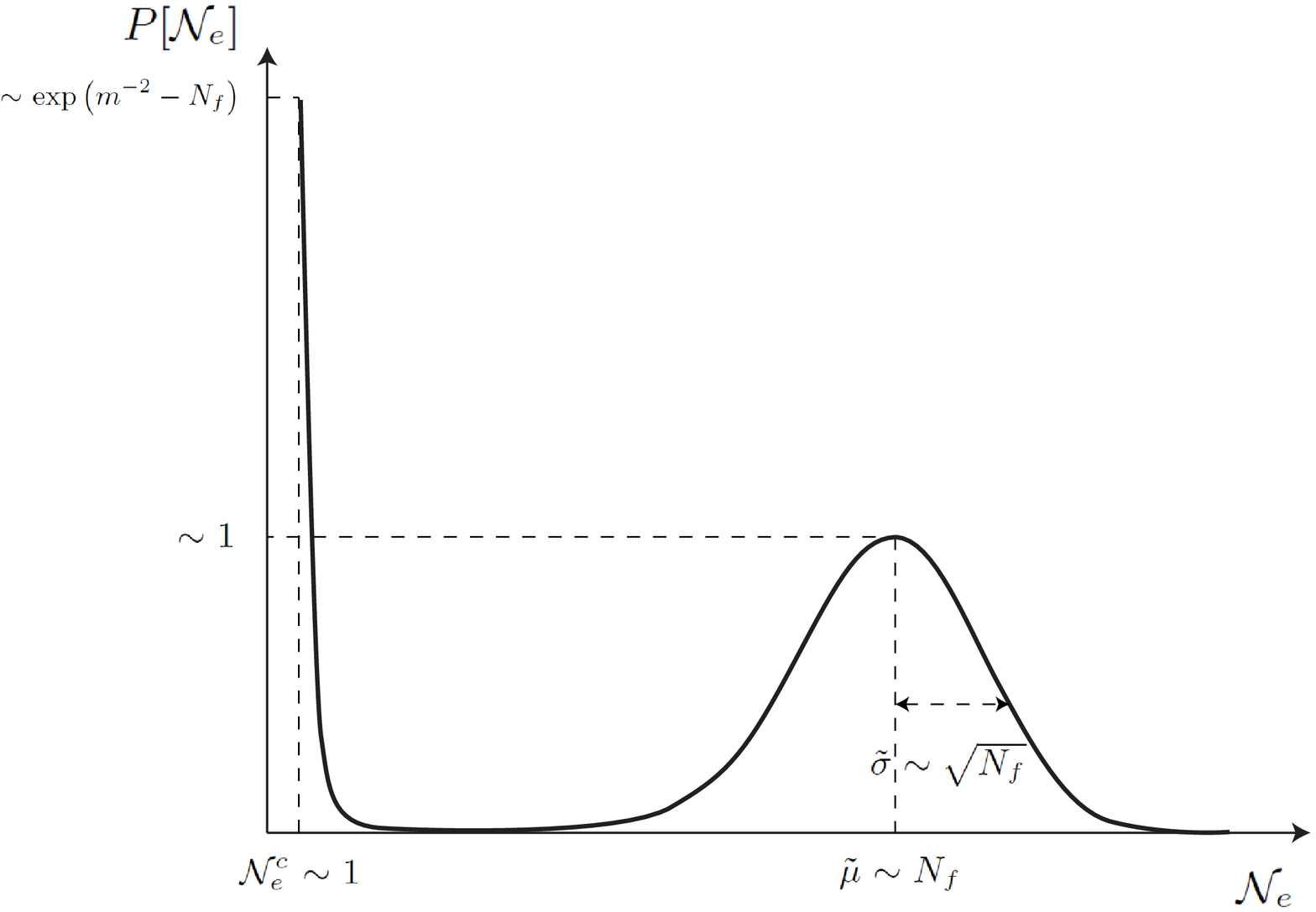}
\caption{\label{schematic_diag}Schematic diagram of the no-boundary measure in Equation~\eqref{nb_largeNf}. For initial fields below the Planck scale, the degeneracy factor of the $e$-folding number has expectation value of the order of $N_{f}$ while its standard deviation is order $\sqrt{N_{f}}$. When $N_{f} \gtrsim m^{-2}$, the peak at $\mathcal{N}_{e}^{c}$ which comes from the Euclidean action term becomes comparable to the peak at $\tilde{\mu}$ which comes from the degeneracy of the field space. Therefore, the degeneracy factor enhances the likelihood of the larger $e$-folding number. Note that the probability in vertical axis is not normalized and intended to represent the relative probability.}
\end{center}
\end{figure}

This distribution is confirmed by the simulation in Figure~\ref{formula_vs_sim}.
The no-boundary measure also can be denoted by the $e$-folding number:
\begin{eqnarray}
\begin{aligned}
dP[\mathcal{N}_{e}] &\varpropto \exp \left(\frac{3 \pi}{2} \frac{1}{m^{2} \mathcal{N}_{e}}
- \frac{\left(\mathcal{N}_{e} - \tilde{\mu}\right)^{2}}{2 \tilde{\sigma}^{2}}\right)\, d\mathcal{N}_e \\
&=\exp \left(\frac{3 \pi}{2} \frac{1}{m^{2} \mathcal{N}_{e}}
- \frac{5}{8} N_{f} \left(1 - \frac{\mathcal{N}_{e}}{\tilde{\mu}}\right)^{2}\right)\, d\mathcal{N}_e. \label{nb_largeNf}
\end{aligned}
\end{eqnarray}
The degeneracy factor enhances the probability at $\mathcal{N}_{e} = \tilde{\mu}$ and hence it increases the likelihood of the larger $e$-folding number. This effect becomes stronger as $N_{f}$ increases and eventually overcomes the probability peak at $\mathcal{N}_{e}^{c}$ when
\begin{eqnarray}
N_{f} \gtrsim m^{-2}, \label{cri_Nf}
\end{eqnarray}
since the parenthesis in the exponential of Equation~\eqref{nb_largeNf} is less than one for $\mathcal{N}_{e} < \tilde{\mu}$. Of course, in the $N_{f} \gg m^{-2}$ limit, the degeneracy term dominates over the Euclidean action term. Figure~\ref{schematic_diag} summarizes the main result -- competition between two contributions -- of this paper\footnote{This conclusion does not sensitively depend on the choice of the cutoff structure. For example, if we do not introduce the cutoff, then we should add the phase volume $\Pi^{N_f}_{i=1} d\phi_{i}$ when we count the probability as a function of $e$-foldings, since the number of $e$-foldings is the function that only depends on the field modulus $\varphi$. The phase volume is
\begin{eqnarray}
\Pi^{N_{f}}_{i=1} d\phi_{i} = \varphi^{N_{f}-1} d\varphi d\Omega_{N_{f}-1}
\end{eqnarray}
and hence
\begin{eqnarray}
dP\left[ \varphi \right] \simeq \exp \left[ \frac{3}{4m^{2} \varphi^{2}} + \left(N_{f}-1\right) \log \varphi \right] d\varphi.
\end{eqnarray}
In the exponent, two factors compete with each other. If $N_{f} \simeq m^{-2}$, then the contribution near the cutoff and the up-lift effect due to the phase volume become similar orders. Hence, we come back to the relation $N_{f} \simeq m^{-2}$. Of course, since we change the cutoff structure, the mathematical details are changed; but this preserves important nature.}.

\subsection{Cosmological viability}
\label{sec_nb_ourU}
%
%
%\subsection{Multi-field inflation}
As previously mentioned, we take $N$-flation models in the same mass case~\cite{DKMW,KL061} as possible candidates of multi-field inflation and apply the results of the no-boundary measure.

Using the slow-roll approximation and for quadratic potentials, the number of $e$-foldings Equation~(\ref{Ne}) can be rewritten as~\cite{Nstar}
\begin{equation}
\mathcal{N}_e \simeq 2\pi\sum \phi_i^2 \,.
\end{equation}
We will assume throughout that the observable scales crossed outside the horizon $60$ $e$-foldings before the end of inflation $\mathcal{N}_{e}^{*} \sim 60$~\cite{Nstar}.

The power spectrum, the spectral index, the tensor-to-scalar ratio, and the non-Gaussianity parameter are given by~\cite{KL061}
\begin{eqnarray}
\allowdisplaybreaks
P_{\mathcal{R}}
	&\simeq& \frac{128 \pi}{3} V(\{\phi_{j}^{*}\}) \sum_{i=1}^{N_{f}} \left(\frac{V(\phi_{i}^{*})}{V(\phi_{i}^{*})_{,\phi{i}}}\right)^2
	\simeq \frac{16 \pi}{3} m^{2} \left(\sum_{i}(\phi_{i}^{*})^{2}\right)^{2}\,,\\
n_{{\rm S}}-1
	&\simeq&  -\frac{1}{\pi \sum_i \phi_i^2  } \,\,
	\simeq -\frac{2}{\mathcal{N}_e^*} \,,\\
r
	&\simeq& \frac{4}{\pi\sum_i \phi_i^2} \,\,
	\simeq \frac{8}{\mathcal{N}_e^*}  \,,\\
\frac{6}{5} \, f_{{\rm NL}}
	&\simeq& \frac{1}{4\pi\sum_i \phi_i^2}
	\simeq \frac{1}{2\mathcal{N}_e^*} \,.
\end{eqnarray}
For $\mathcal{N}_{e}^{*} \sim 60$~\cite{Nstar}, these become
\begin{eqnarray}
n_{{\rm S}}-1 & \sim & 0.033 \,,\\
r & \sim& 0.133 \,,\\
\frac{6}{5} \, f_{{\rm NL}}&\sim& O(0.01) \,.
\end{eqnarray}
Even though this case does not give significantly large non-Gaussianity, but still all these parameters are still within viable region of WMAP7~\cite{WMAP7}.

We can reduce the formula for the power spectrum in terms of effect single field as
\begin{eqnarray}
P_{\mathcal{R}} \simeq %\frac{16 \pi}{3} m^{2} \left(\sum_{i=1}^{N_{f}}(\phi_{i}^{*})^{2}\right)^{2}
%=
\frac{16 \pi}{3} m^{2} (\phi^{*})^{4}.
\end{eqnarray}The observed normalization of the spectrum can be taken as $P_{\mathcal{R}}^{1/2} \simeq 5 \times 10^{-5}$~\cite{WMAP7}, leading to a normalization of
\begin{eqnarray}
m \sim 1.3 \times 10^{-6}, \label{obs_m}
\end{eqnarray}
which is a reasonable mass scale for the quadratic model regardless of the choice of $N_{f}$.
Therefore, from Equation~\eqref{cri_Nf}, the multi-field factor works for $N_{f} \gtrsim 10^{12}$ which is quite large number of fields needed\footnote{Note that the volume weighting scenario \cite{Hartle:2007gi,Hartle:2008ng} also suffers a similar problem. The condition $N_{f} \sim m^{-2}$ is comparable to that of the volume factor in Equation~\eqref{cri_Ne}, because when the cutoff $\alpha$ is order of the Planck scale, a set of fields satisfying Equation~\eqref{cri_Nf} will produce enough $e$-folding number to satisfy Equation~\eqref{cri_Ne}. Hence, for a single field quadratic potential model, to enhance sufficient $e$-foldings using the volume weighting, we need to allow the field space up to an extremely huge value $\varphi \sim 10^{5}$.}.
\section{Discussions and conclusion}
\label{sec_Conclusion}
In this article, we investigated the no-boundary measure with $N_f$ fields. To obtain a very simple model, we assumed a quadratic potential with a same mass $m$. Following the methods of $N$-flation, we further assumed that the field values of each fields should be less than the Planck scale: $\alpha \lesssim 1$. Then, the no-boundary measure effectively has two competing factors (Equation~(\ref{nb_multi_phisq})): the Euclidean action and the Gaussian degeneracy factor.

If the number of fields $N_{f}$ becomes sufficiently large so that $N_{f} \sim m^{-2}$, then the two factors are of similar orders. The peak of the Gaussian factor $\varphi \sim N_{f}$ then gives realistic cosmologies non/negligible probability.  However, the number of fields needed to achieve this effect seems to be very large. In the simple model we considered, $N_{f} \sim 10^{12}$ number of fields are needed to fit the power spectrum of our universe. Since this is a very high number, this seems to suggest that this is not an effective way to resolve the tension between no-boundary probabilities and the observations. But there are also less pessimistic ways to view the situation:
\begin{enumerate}
\item A large number of scalar fields is not ruled out by experiments if their energy scale is near Planck scale. Indeed it has been argued that such a scenario is useful to resolve some problems of our phenomenological universe \cite{Dvali:2007hz}. Therefore, it may still be a viable idea.
\item In our calculation, we assumed a \emph{quadratic potential} described by a \emph{single constant mass}, but one can certainly question these assumptions.
\begin{enumerate}
\item The masses can depend on the energy scale or the time, via the running of couplings (masses) or the temperature dependence. If the effective mass is sufficiently large when the universe begins inflation, and if it decreases as time goes on, then one may explain the sufficient $e$-foldings with reasonable numbers of fields.
\item The mass can be different for different fields, e.g., Refs.~\cite{KL061,RM}. If the Euclidean dynamics highly depends on the larger mass, then the required number of fields to lift up the larger $e$-foldings can be reduced. This can explain the power spectrum, if the smallest mass is order of $10^{-6}$ \cite{Lyth:2001nq}.
\item The quadratic potential should only be regarded as a first approximation of more complicated potentials (e.g., axions in Refs.~\cite{KLS}). More complicated potentials may change the Euclidean cutoffs and reduce the peak of the Euclidean probability, leading to a required $N_f$ more in keeping with present models of high energy physics.
\item If the fields couple in complicated ways, then again, the Euclidean cutoff structures can be changed. Therefore, from the similar arguments, one may reduce the number of fields to a reasonable amount.
\end{enumerate}
\end{enumerate}
To summarize, we have shown that considering the no-boundary wave function for many fields can solve the traditional problem of low probabilities for sufficient inflation. But in the simple model we have considered, the number of fields is very high, much higher than supported by, for example, string theory scenarios. Further research is needed to see whether the high number of necessary fields can be brought down in more complicated and realistic models.

\paragraph{Note for recent progress} Recently, Planck data \cite{Ade:2013uln} was revealed and the naive model with the $\sim \phi^{2}$ potential seemed not preferred by the data. Therefore, to build a better model using multi-field inflation, one may need to include the curvaton \cite{Lyth:2001nq}. In any case, our main idea still survives and needs further investigations. We remain for future work. For recent discussion in the context of Planck, see \cite{Kaiser:2013sna,Easther:2013bga}.

\section*{Acknowledgment}

We would like to thank Robert Brandenberger and Ewan Stewart for helpful discussions. DH, DY and BHL were supported by the National Research Foundation of Korea(NRF) grant funded by the Korea government(MEST) through the Center for Quantum Spacetime(CQUeST) of Sogang University with the grant number 2005-0049409. DY is supported by the JSPS Grant-in-Aid for Scientific Research (A) No.~21244033. SAK was supported by Basic Science Research Program through the National Research Foundation of Korea(NRF) grant funded by the Korea government(MEST) with the grant number 2011-0011083. SAK and DY would like to thank the organizers of the 2012 Asia Pacific School/Workshop on Cosmology and Gravitation and the KASI-APCTP joint mini-workshop on Science and Technology of BigBoSS. SAK thanks to the hospitality of Laila Alabidi, Takahiro Tanaka and YITP, and of CQUeST while this work was being carried out.

%\newpage

\end{document}